\documentclass[12pt]{article}
\usepackage{epsfig}
\usepackage{amsmath}
\usepackage{hhline}
\usepackage{amssymb}
\usepackage{times}
\usepackage{color}
\usepackage{cite}

\newlength{\dinwidth}
\newlength{\dinmargin}
\begin{document}

\newcommand {\gapprox}
   {\raisebox{-0.7ex}{$\stackrel {\textstyle>}{\sim}$}}
\newcommand {\lapprox}
   {\raisebox{-0.7ex}{$\stackrel {\textstyle<}{\sim}$}}
\def\gsim{\,\lower.25ex\hbox{$\scriptstyle\sim$}\kern-1.30ex%
\raise 0.55ex\hbox{$\scriptstyle >$}\,}
\def\lsim{\,\lower.25ex\hbox{$\scriptstyle\sim$}\kern-1.30ex%
\raise 0.55ex\hbox{$\scriptstyle <$}\,}

%
%

\begin{titlepage}

\noindent
\begin{flushleft}
{\tt DESY 09-099    \hfill    ISSN 0418-9833} \\
{\tt June 2009}                  \\
\end{flushleft}


\vspace{1.0cm}

\begin{center}
\begin{Large}

{\bf Chargino and Neutralino Separation with the ILD Experiment} 

\vspace{1.5cm}

T.~Suehara$^{1}$, J.~List$^{2}$

\end{Large}

\vspace{.3cm}
1- International Center for Elementary Particle Physics\\ 
     The University of Tokyo, Hongo 7-3-1, Bunkyo District\\
     Tokyo 113-0033, Japan
\vspace{.1cm}\\
2- Deutsches Elektronen Synchrotron DESY\\ 
   Notkestr. 85 \\
D-22607 Hamburg, Germany 
     
\end{center}

\vspace{1cm}

\begin{abstract}
One of the benchmark processes for the optimisation of the detector concepts 
proposed for the International Linear Collider is Chargino 
and Neutralino pair 
production in an mSugra scenario where $\tilde{\chi}^{\pm}_1$ and 
$\tilde{\chi}^0_2$ are mass degenerate and decay into 
$W^{\pm}\tilde{\chi}^0_1$ and $Z^0\tilde{\chi}^0_1$, respectively. 
In this case the separation of both processes in the fully hadronic 
decay mode is very sensitive to the jet energy resolution and thus 
to the particle flow performance.
The mass resolutions and cross-section uncertainties achievable with the ILD 
detector concept are studied in full simulation at a center of mass 
energy of 500~GeV, an integrated luminosity of 500~fb$^{-1}$ and beam 
polarisations of  $P(e^+,e^-) = (30\%, -80\%)$.
For the $\tilde{\chi}^{\pm}_1$ and  $\tilde{\chi}^0_2$ pair production 
cross-sections, statistical precisions of 0.84\% and 2.75\% are 
achieved, respectively. The masses of $\tilde{\chi}^{\pm}_1$, 
$\tilde{\chi}^0_2$ and $\tilde{\chi}^{0}_1$ can be determined 
with a statistical precision of 2.9 GeV, 1.7 GeV and 1.0 GeV, 
respectively.
\end{abstract}

\vspace{1.0cm}

\begin{center}
Submitted to {\em Eur. Phys. J.} {\bf C}
\end{center}

\end{titlepage}

\newpage

\section{Introduction}
\label{sect:intro}

In anticipation of the International Linear Collider (ILC), a proposed $e^+e^-$ 
collider with center-of-mass energies between 90 and 500~GeV, upgradable to 1~TeV, 
and polarised beams, 
several detector concepts are being discussed. In order to evaluate the performance of 
these concepts, benchmark processes have been chosen which 
are challenging for key aspects of the detector designs~\cite{ILCbenchmarks}.

In order to test the jet energy resolution, a supersymmetric scenario
which assumes non-universal soft SUSY-breaking contributions to the 
Higgs masses has been defined. In this scenario, the mass differences between 
the lightest SUSY particle (LSP) and the 
heavier gauginos become large, while at the same time the sleptons 
are so heavy that gaugino decays into sleptons are kinematically 
forbidden. The corresponding benchmark point has been defined 
in~\cite{ILCbenchmarks} as ``Point 5'' with the following SUSY parameters:

\begin{equation}
m_0 = 206~\mathrm{GeV,} \quad m_{1/2} = 293~\mathrm{GeV,} \quad \tan{\beta} = 10, \quad
 A = 0, \quad \mu=375~\mathrm{GeV} 
\end{equation} 

With a top quark mass of $M_t = 178$~GeV, the following gaugino masses 
are obtained by Spheno~\cite{spheno}:

\begin{equation}
M_{\tilde{\chi}^0_1} = 115.7~\mathrm{GeV, } \quad M_{\tilde{\chi}^{\pm}_1} = 216.5~\mathrm{GeV, } \quad  M_{\tilde{\chi}^0_2} = 216.7~\mathrm{GeV, } \quad M_{\tilde{\chi}^0_3} = 380~\mathrm{GeV}. 
\end{equation} 

The lightest sleptons are even heavier than the gauginos, thus leading 
to branching fractions of 99.4\% for the decay  $\tilde{\chi}^{\pm}_1 
\rightarrow W^{\pm} \tilde{\chi}^0_1$ and 96.4\% for 
$\tilde{\chi}^{0}_2 \rightarrow Z^0 \tilde{\chi}^0_1$:

\begin{equation}
M_{\tilde{\tau}_1} = 230.8~\mathrm{GeV} \qquad M_{\tilde{e}_R} = 237.4~\mathrm{GeV} 
 \end{equation}

In order to benchmark the jet energy reconstruction, the 
fully hadronic decay mode of the gauge bosons is considered here. 
In this mode, Chargino and Neutralino events can only be separated 
via the mass of the vector bosons they decay into. The motivation 
of this study is not to evaluate the final precision which could be 
achieved at the ILC by combining several final states, or even by 
performing threshold scans, but to 
test the detector performance in the most challenging decay mode.

The analysis is performed at a center of mass energy of 500~GeV for 
an integrated luminosity of 500~fb$^{-1}$ with beam polarisations 
of $P(e^+,e^-) = (30\%, -80\%)$. It is based on a detailed simulation 
of the ILD detector based on GEANT4~\cite{geant4}, which is described briefly in the next section.
Section~\ref{sect:reco} discusses the event reconstruction and selection 
procedure, including a pure Standard Model control selection. The results for the cross-section and mass measurement are presented in sections~\ref{sect:xsection} and \ref{sect:mass}, respectively.

 

\section{The ILD Detector Concept and its Simulation}
\label{sect:ILD}
The proposed ILD detector has been described in detail in the ILD Letter of Intent~\cite{ILD}. Its main characteristics comprise a time projection chamber as
a main tracking device, which is complemented by silicon tracking and vertexing detectors, and highly granular electromagnetic and hadronic calorimeters as required for the particle flow approach~\cite{pandora}. Both, tracking system and calorimeters, are included in a solenoidal magnetic field with a strength of 3.5~T 
provided by a superconducting coil. The magnetic flux is returned in an iron yoke,
which is instrumented for muon detection. Special calorimeters at low polar angles
complement the hermeticity of the detector and provide luminosity measurement.

While previous studies were based on fast simulation programs which smear four-vectors with expected resolutions, we have used a full GEANT4 based simulation of all ILD 
compoments. Many details are included, in particular gaps in the sensitive regions and realistic estimates of dead material due to cables, mechanical support,
cooling and so on.

With this detector simulation, the following performance has been achieved~\cite{ILD}: For tracks with a transverse momentum $p_t$ larger than 1~GeV, the tracking efficiency is 99.5\% across almost the entire polar angle range of $|\cos{\theta}|<0.995$ covered by the tracking detectors, with a $p_t$ resolution of better than $\sigma_{1/p_t} = 2 \times 10^{-5} \oplus 1 \times 10^{-3} / (p_t \sin{\theta})$. The calorimetric system has been designed to deliver a jet energy resolution of 3.0\% to 3.7\% over a large range of energies from 250 GeV down to 45 GeV for polar angles $\theta$ in the range $|\cos{\theta}| < 0.9$. The luminosity is expected to be known to $10^{-3}$ from measurements of the Bhabha scattering cross-sections at small angles. The beam polarisations and the beam energies will be measured to $\delta P/P = 0.25\%$ and $2 \times 10^{-4}$, respectively by dedicated instrumentation in the beam delivery system.  

The event sample used in this analysis has been generated using the matrix element generator Whizard~\cite{whizard}. It comprises all Standard Model processes plus
all kinematically accessible SUSY processes in the chosen scenario. In total, about $12 \times 10^6$
events have been generated and processed through the full simulation and reconstruction chain for this analysis.
 

\section{Event Reconstruction and Selection}
\label{sect:reco}
The reconstruction and also the first event selection steps are implemented in the 
MarlinReco framework~\cite{marlin}. The central part of the reconstruction for this 
analysis is the particle flow algorithm Pandora~\cite{pandora}, which forms charged 
and neutral particle candidates - so-called ``particle flow objects'' or PFOs - from 
tracks and calorimeter clusters. The resulting list of PFOs for each event is forced 
into a 4--jet configuration using the Durham algorithm. The jet energy scale is 
raised by 1\%, determined from dijet samples. No special treatment of b-quark jets 
is considered here.

As a final step of the reconstruction, a constrained kinematic fit~\cite{kinfitnote}, which 
requires the two dijet masses of the event to be equal, is performed on each event. All three possible jet pairings are tested. The resulting improvement in mass resolution is evaluated
on Standard Model events, as described in section~\ref{sec:smcontrol}.


%
%
\subsection{SUSY Selection}
\label{sec:presel}

The major part of the Standard Model events is rejected by applying the
following selection to all events in the SUSY and SM samples:

\begin{itemize}
  \item In order to eliminate pure leptonic events, the total number 
  of tracks in the event should be larger than $20$ and each jet has 
  to contain at least two tracks.
  \item Since the two LSPs escape undetected, the visible energy of the 
  event $E_{\mathrm{vis}}$ should be less than 300~GeV. In order to
  remove a substantial fraction of 2-photon events with very low 
  visible energy, $E_{\mathrm{vis}}> 100$~GeV is required as well.
  \item To ensure a proper jet reconstruction, each jet should have
  a reconstructed energy of at least 5~GeV and a polar angle $\theta$ 
  fulfilling $|\cos(\theta_{\mathrm{jet}})| < 0.99$.
  \item 2-jet events are rejected by requiring the distance parameter
  of the Durham jet algorithm for which the event flips from 4-jet to 3-jet
  configuration, $y_{34}$ to be larger than 0.001.
  \item Coplanar events (e.g.~$W^+W^-$ with ISR/beamstrahlung photons) 
  are removed by requiring $|\cos(\theta)|$ of the missing momentum \
  to be smaller than 0.99.
 \item No lepton candidate with an energy larger than 25~GeV is allowed
  in order to suppress semi-leptonic events. 
\end{itemize}

The upper part of table~\ref{tbl:cutstat} shows the reduction for these cuts.
The selection efficiency of hadronic Chargino and Neutralino pair events is 
very high, 88.1\% and 90.8\%, respectively. Therefore, we will refer to this stage 
in the selection process as ``high efficiency'' selection. Although the SM background 
is significantly reduced already by these cuts, the contribution from 4-fermion events 
is still large, about 6 times the Chargino signal. 

\begin{figure}[p] 
\begin{center}
\epsfig{figure=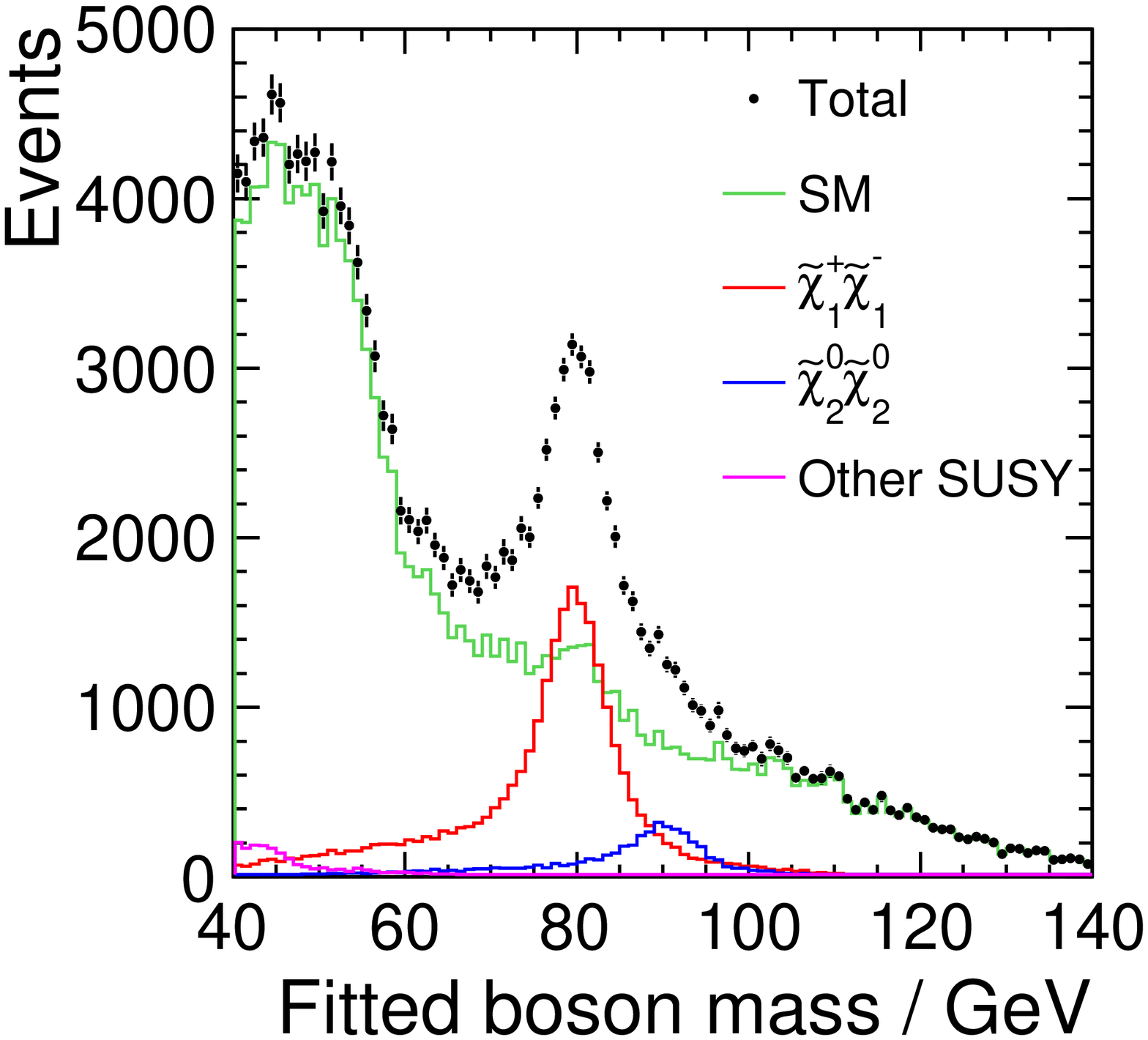,width=7cm}
\vspace{0.001cm}
\epsfig{figure=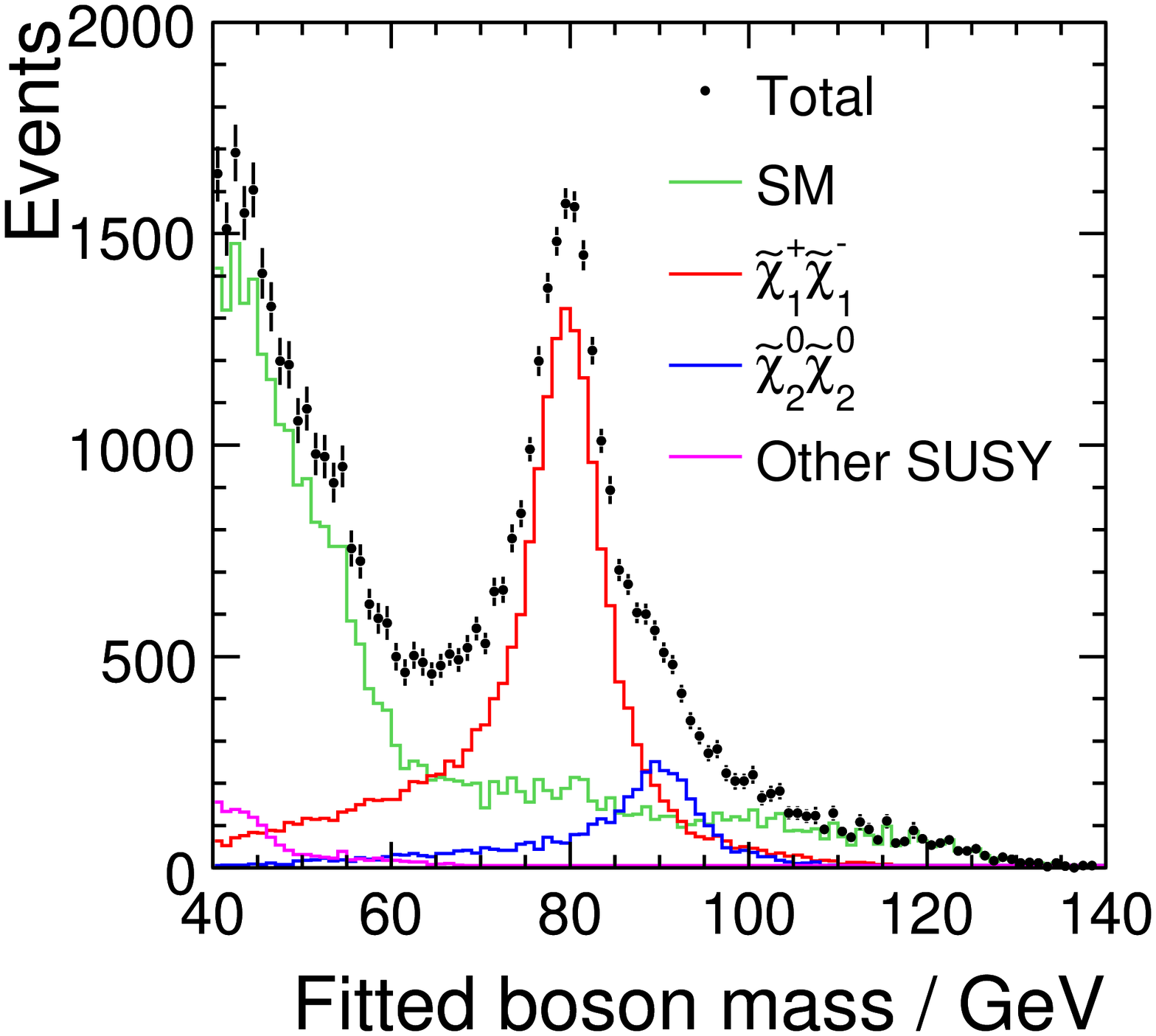,width=7cm}
\put(-115.,55.){a)}
\put(-45.,55.){b)}
\end{center}
  \caption{\label{fig:fitmass_allcuts} a) Reconstructed mass of the vector boson candidates after all selection cuts and kinematic fit for the jet pairing with the highest fit probability. b) Same distribution after some additional cuts to enhance the purity.}
\end{figure}

Figure~\ref{fig:fitmass_allcuts}a) shows the reconstructed boson mass 
distribution as obtained by the constrained kinematic fit after these 
selection cuts. A large fraction of the remaining Standard Model background 
features low invariant dijet masses, but nevertheless a 
sizable amount of background remains also in the signal region.

For the cross-section measurement, the sample is therefore cleaned further 
by four additional cuts: 

\begin{itemize}
  \item The number of particle flow objects (PFOs) in each jet should be $N_{\mathrm{PFO}}> 3$ 
        in order to reject $\tau$ jets more effectively.
  \item The direction of the missing momentum should fulfill $|\cos{\theta_{\mathrm{pmiss}}}| < 0.8$:
        This cut is quite powerful to reject all kinds of SM backgrounds, which tend to 
        peak in the forward region, while the signal follows a flat $\cos{\theta_{\mathrm{pmiss}}}$
        distribution. 
        Nevertheless, it reduces the signal efficiency substantially, which could be avoided for 
        example by placing a more stringent cut on the missing mass instead (see next item). 
        However, the missing mass distribution of the signal directly depends on the LSP mass, 
        thus it should not be too finely tuned to specific mass values, since we want to measure 
        the gaugino masses. The prediction of a flat $\cos{\theta_{\mathrm{pmiss}}}$ distribution 
        depends only on the spin, and can thus be considered model-independent (within SUSY).
  \item The missing mass should be larger than 220~GeV to further reject 6-fermion events 
        (semi-leptonic $t\bar{t}$). The value of this cut is chosen such that it is in a region 
        with no SUSY contribution, i.e. where the data should agree with the SM expectation. 
        Thus in a real experiment an adequate cut position could be found from the data. 
        For this reason, no upper cut is placed on $M_{\mathrm{miss}}$, since other SUSY 
        processes contribute there, and it would not be trivial to determine a suitable cut 
        value from real data.
  \item The kinematic fit constraining the two dijet masses to be equal should converge for at 
        least one jet pairing: This is necessary in order to use the fit result for further
        analysis. The efficiency and resolution of the fit can be cross-checked easily on 
        real data, for instance with the control selection decribed in the previous section.
\end{itemize} 

The obtained reduction due to these cuts is shown in the last four lines of 
table~\ref{tbl:cutstat}.
The final distribution of the reconstructed boson mass, again obtained by the constrained 
kinematic fit, is displayed in figure~\ref{fig:fitmass_allcuts}b. It illustrates the 
achieved boson mass 
resolution and thus $W$ and $Z$ pair separation, however at significantly reduced
efficiency. Fitting the total spectrum by a fourth order polynomial for the 
background plus the sum of two Breit-Wigner functions folded with a Gaussian 
for the $W$ and $Z$ contributions, the mass resolutions can be determined to 
3.4~\%. 

Table~\ref{tbl:cutstat2}  shows the final purity and efficiency of signal and 
major background processes. According 
to this table, $e^+e^-\rightarrow qqqq$ is the dominant process in the remaining background.



%
%
\subsection{Standard Model Control Selection}
\label{sec:smcontrol}
\begin{figure}[t] 
\begin{center}
\epsfig{figure=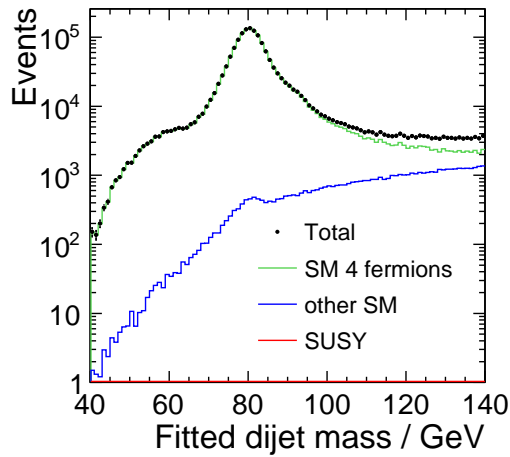,width=7cm}
\end{center}
  \caption{\label{fig:sm_sel} Dijet mass spectrum for Standard Model selection. The event sample is dominated by 4-fermion events, with a small contribution from 6-fermion events, but doesn't contain any SUSY events.}
\end{figure}

Since the Chargino and Neutralino separation relies on reconstructing the masses of the 
$W$ and $Z$ bosons from their decay products, the dijet mass resolution is a crucial parameter 
in this analysis and has to be determined from Standard Model $W$ and $Z$ pair events. For this 
purpose, the ``high efficiency'' selection from above is applied to all simulated data, inverting 
only the cut on the visible energy to $E_{\mathrm{vis}}>300$~GeV. This yields an event sample which
is vastly dominated by 4-fermion events, with a small contribution from 6-fermion events, 
but no SUSY events. The corresponding dijet mass spectrum is shown in figure~\ref{fig:sm_sel}.

The mass resolution has been determined for two cases:
\begin{itemize}
  \item[a)] The jet pairing is chosen such that the difference between the two dijet masses in 
            each event is minimized.
  \item[b)] A kinematic fit, which constrains the two dijet masses in each event to be equal, is 
            performed for all three possible jet-boson associations. The jet pairing which yields 
            the highest fit probability is chosen.
\end{itemize}

The resulting mass distributions are fitted with the sum of two Breit-Wigner functions convoluted
with a Gaussian, fixing the $W$ and $Z$ widths as well as the $Z$ pole mass to their PDG values and
having the same $\sigma$ for both Gaussians, plus a forth order polynomial for all non-resonant contributions.

\begin{figure}[b] 
\begin{center}
\epsfig{figure=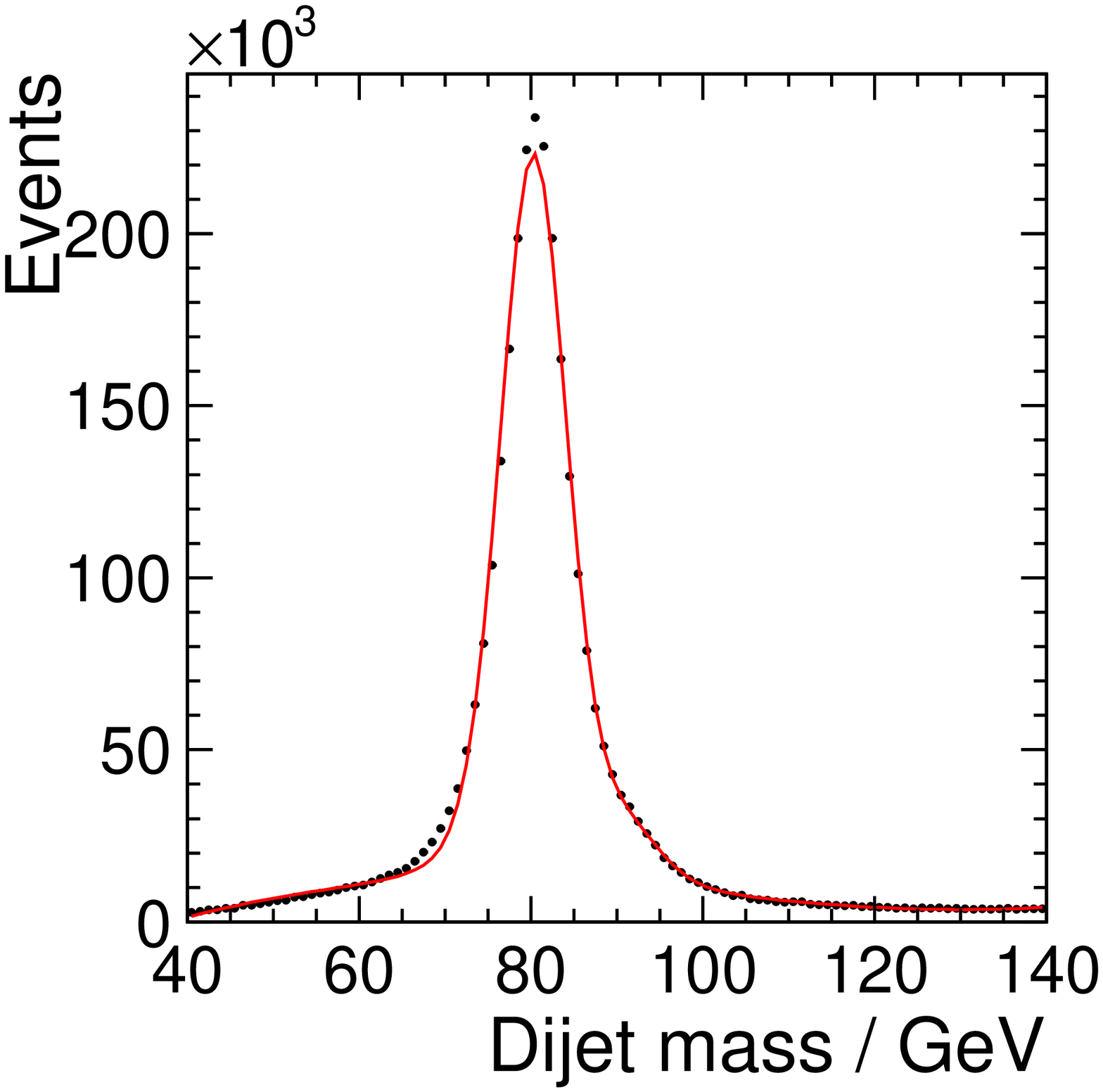,width=7cm}
\vspace{0.001cm}
\epsfig{figure=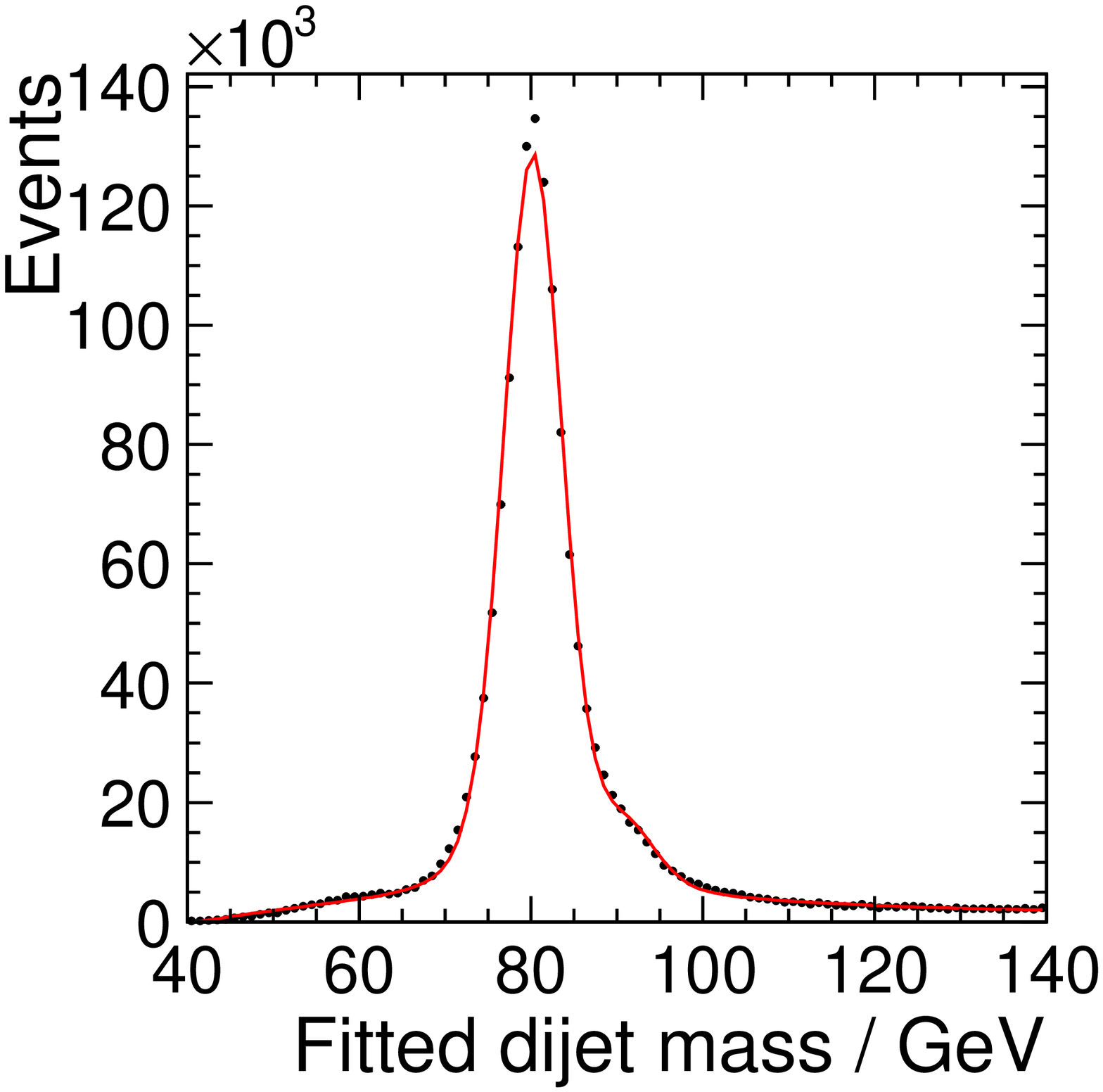,width=7cm}
\put(-120.,55.){a)}
\put(-50.,55.){b)}
\end{center}
  \caption{\label{fig:sm_fits} Dijet mass distributions  a) without and b) with kinematic fit. Fitting the distributions with the sum of two Breit-Wigner functions folded with Gaussian plus a forth order polynomial for the non-resonant background yields dijet mass resolutions of 3.5~GeV (case a) and 3.0~GeV (case b).}
\end{figure}

Figure~\ref{fig:sm_fits} shows the fitted spectra and the resulting fit parameters. In case a), 
without the kinematic fit, the dijet mass resolution is determined as $\sigma_m^a = 3.5$~GeV, 
while it is reduced to $\sigma_m^b = 3.0$~GeV when the kinematic fit is applied.

These mass resolutions are even better than in the SUSY case, since the kinematics of the events is more favourable here. While the SM gauge boson pairs are highly boosted and thus finding the correct jet pairing is relatively easy, the bosons in our SUSY scenario are produced nearly at rest, resulting in a higher combinatorical background and a slightly worse
boson mass resolution. Nevertheless, a SM control selection will be crucial to demonstrate the
level of detector understanding, since the actual SUSY measurement will rely on template distributions and selection efficiencies determined from simulations.

%
%

\section{Cross-Section Measurement}
\label{sect:xsection}

The cross-sections of $e^+e^-\rightarrow\tilde{\chi}^+_1\tilde{\chi}^-_1$ and
$e^+e^-\rightarrow\tilde{\chi}^0_2\tilde{\chi}^0_2$ can be measured by
determining the amount of $W$ and $Z$ pair like events.
For the hadronic events we are concerned with here, a 2-dimensional fit
in the plane of the two dijet masses per event is performed to obtain the 
amount of $W$ and $Z$ pair candidates.

\begin{figure}[p]
	\begin{minipage}{0.47\textwidth}
		\begin{center}
			\includegraphics[width=1\textwidth]{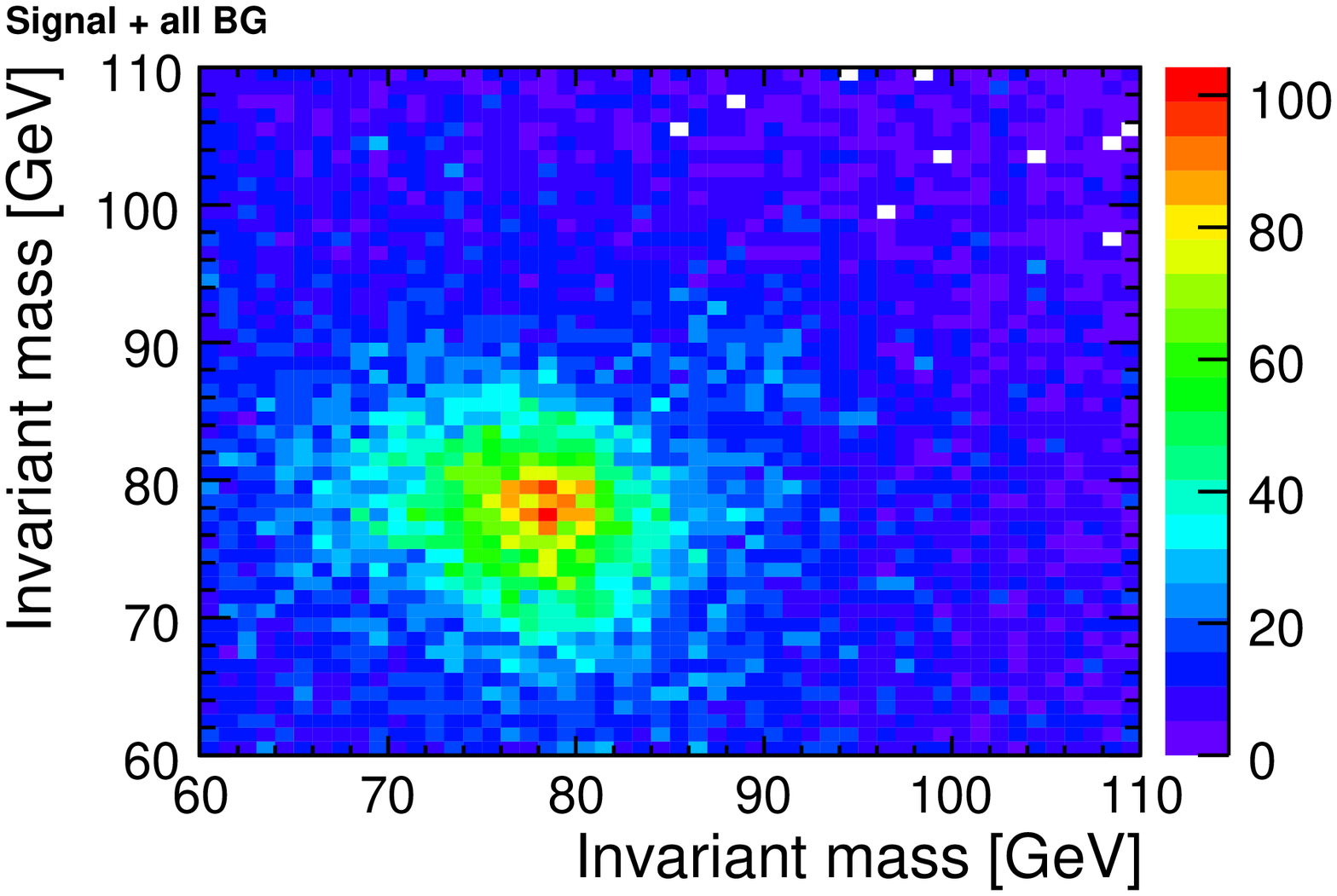}
			\par
				{(a) All events (Sig + BG).}
			\includegraphics[width=1\textwidth]{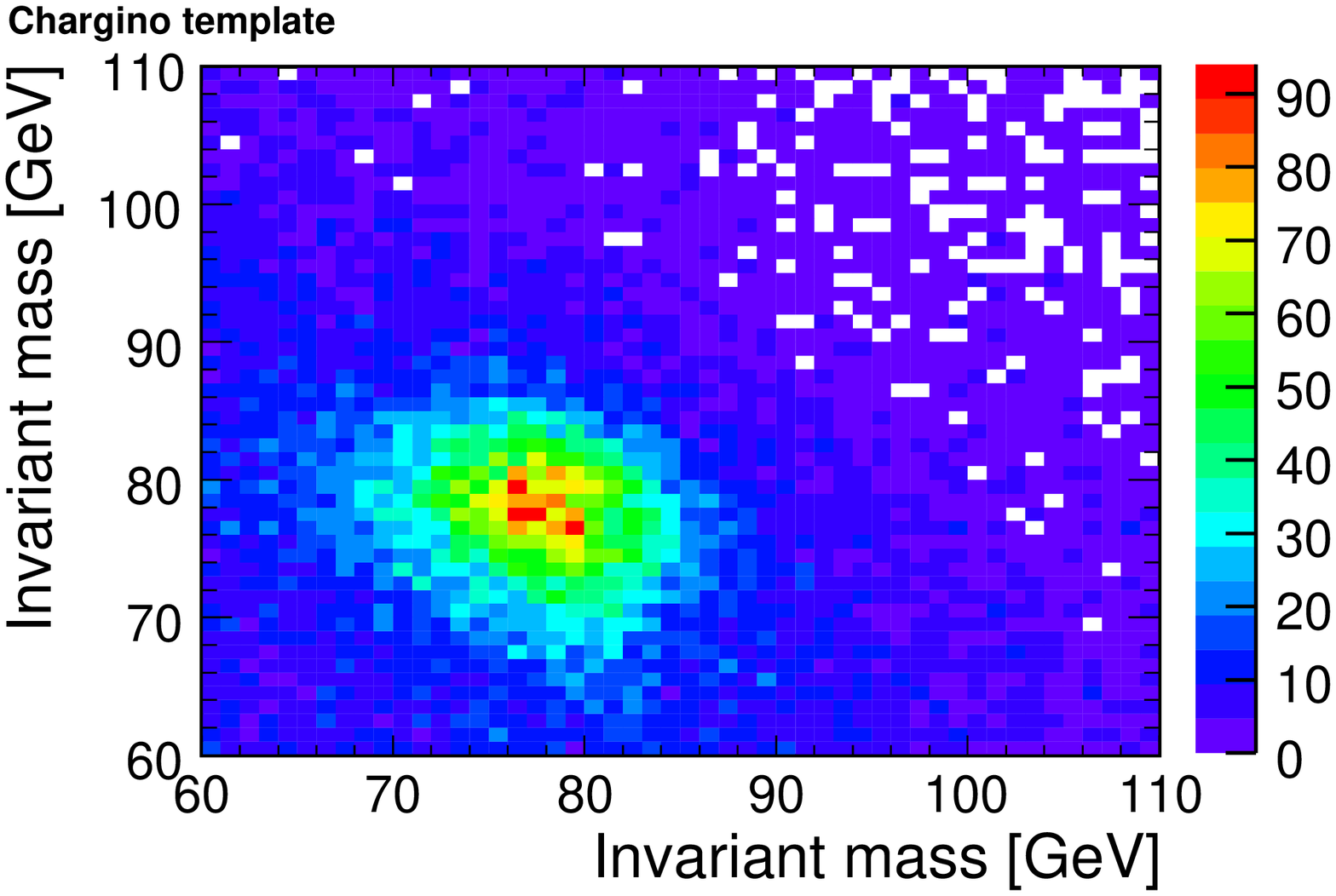}
			\par
				{(c) Chargino-pair template (including all W decay mode).}
			\includegraphics[width=1\textwidth]{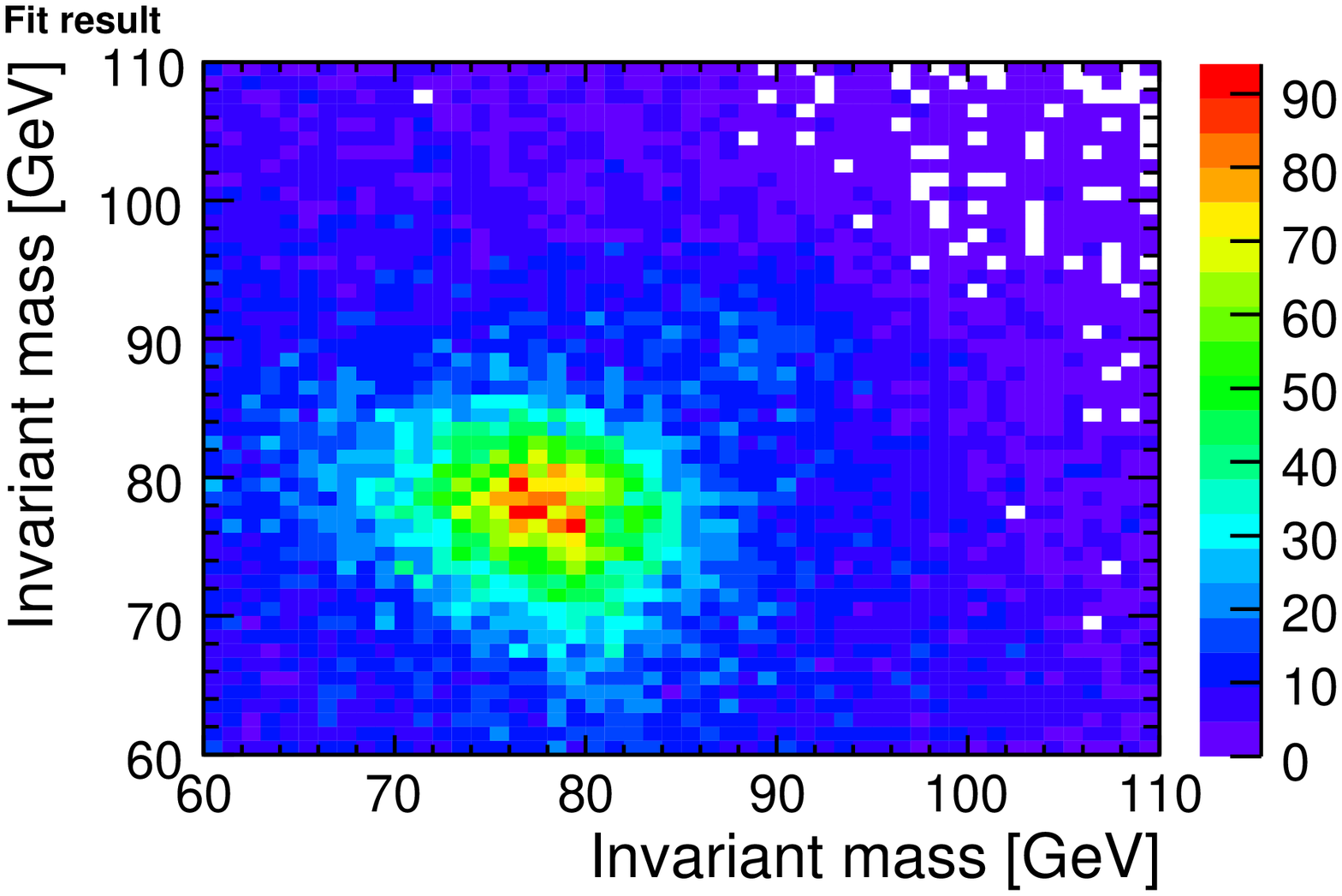}
			\par
				{(e) Fit result to (a) - (b) with SM subtraction fluctuation.}
		\end{center}		
	\end{minipage}
	\begin{minipage}{0.47\textwidth}
		\begin{center}
			\includegraphics[width=1\textwidth]{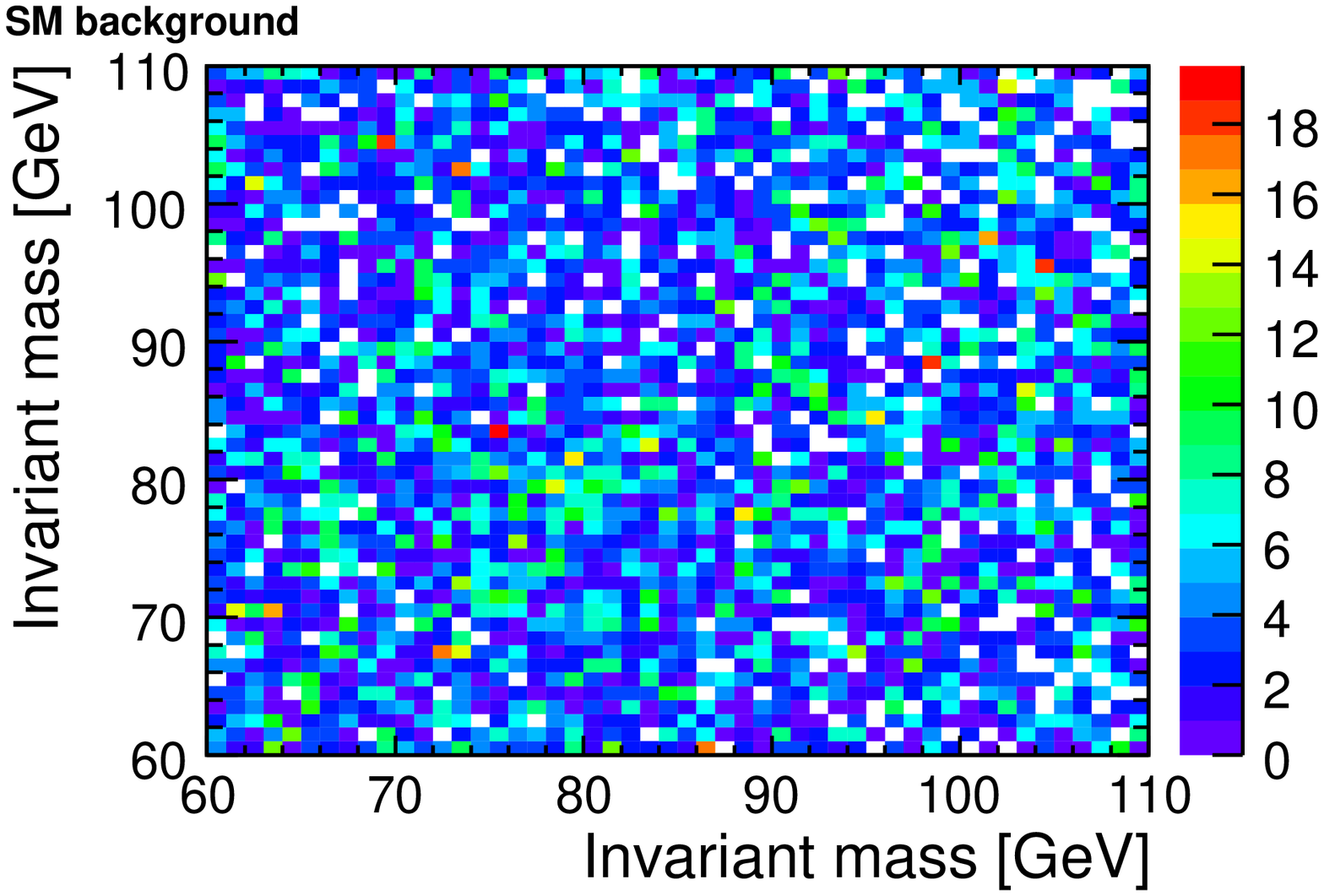}
			\par
				{(b) SM background.}
			\includegraphics[width=1\textwidth]{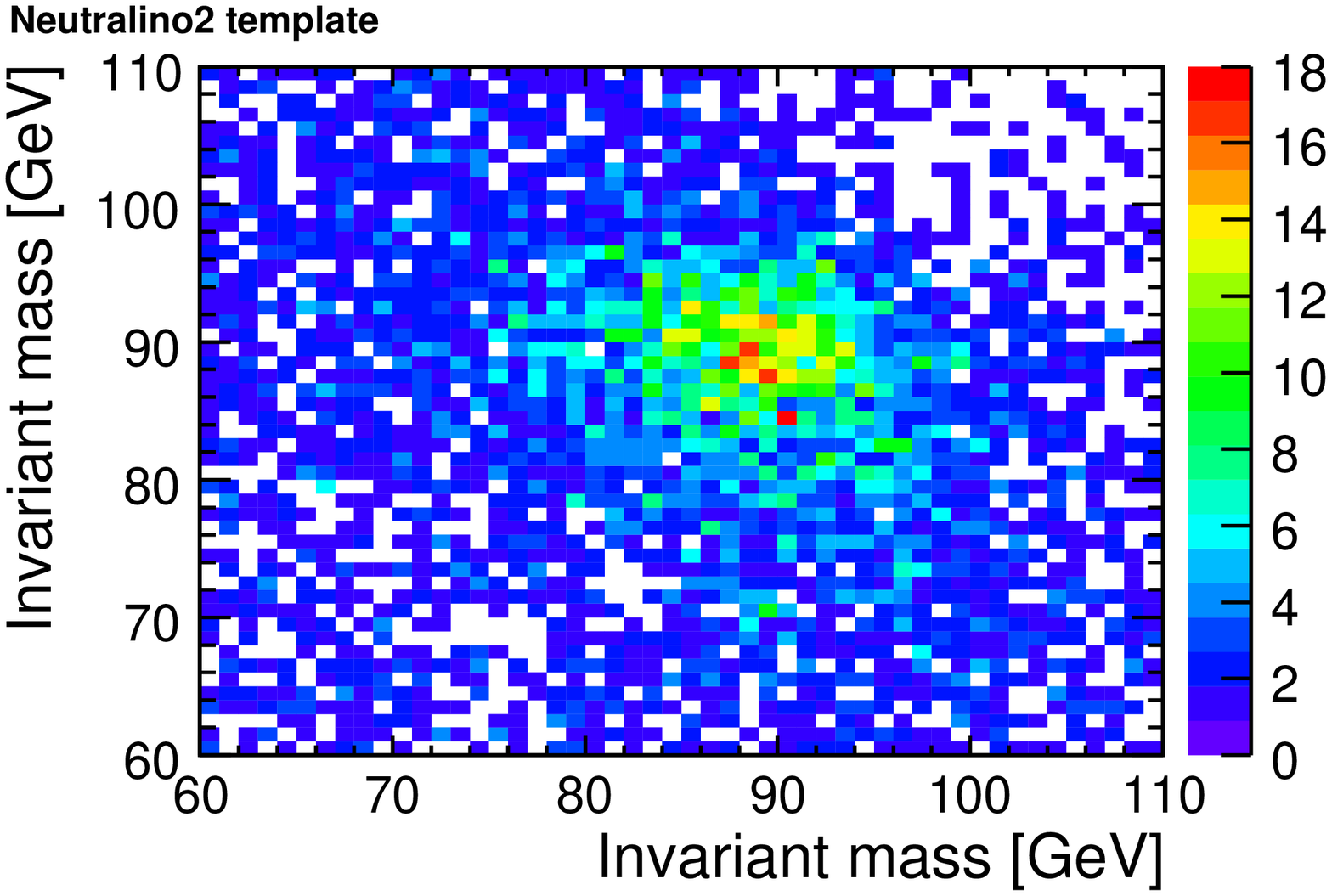}
			\par
				{(d) Neutralino2-pair template (including all Z decay mode).}
			\includegraphics[width=1\textwidth]{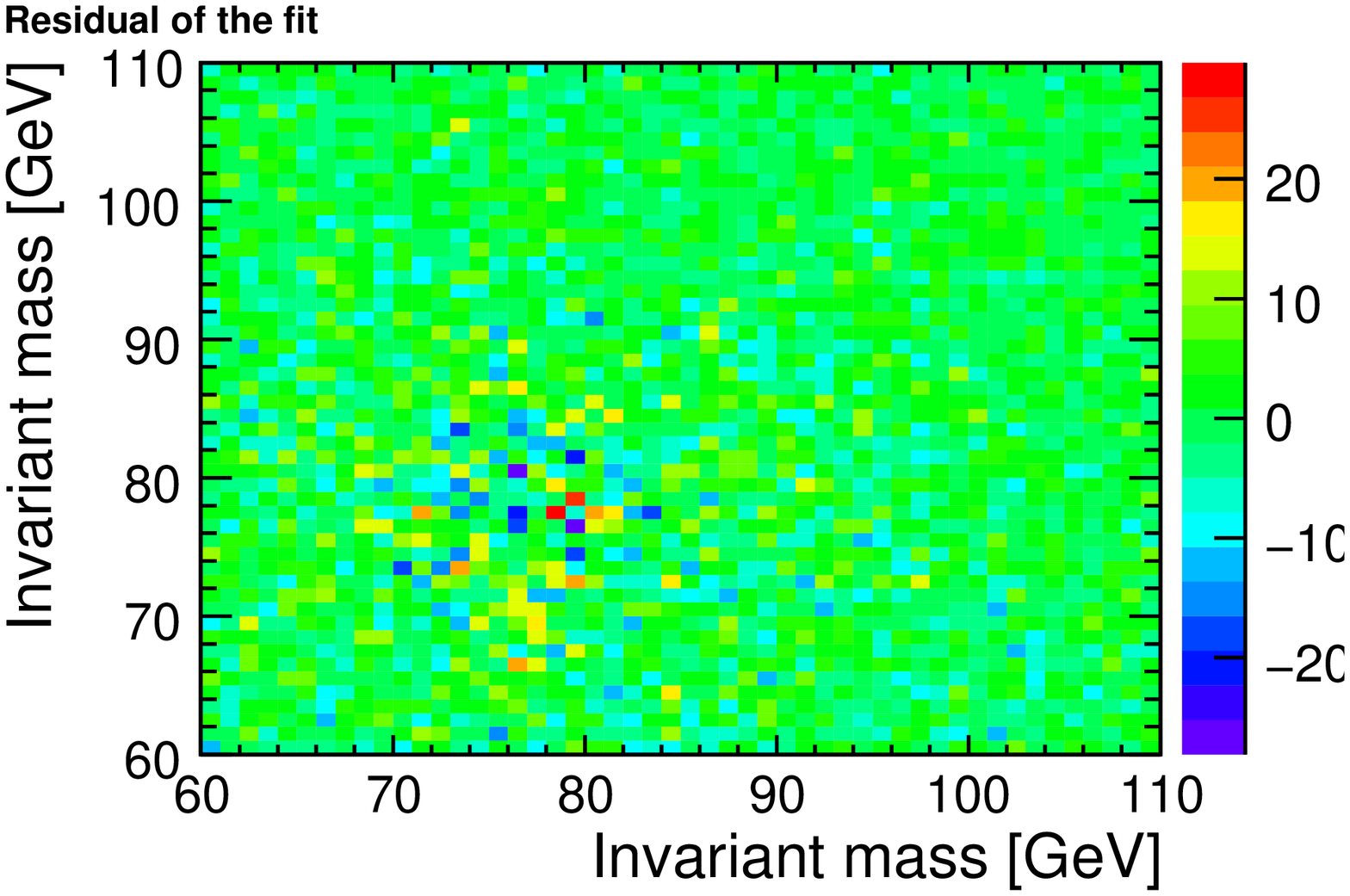}
			\par
				{(f) Residual of the fit: (a) - (b) - (e).}
		\end{center}		
	\end{minipage}
	\caption{Dijet mass distribution for cross-section fit. For (a) and (b) the same events are used,
		while (c) and (d) are statistically independent of (a).}
	\label{fig:djm-csfit}
\end{figure}

Figure~\ref{fig:djm-csfit} shows the dijet mass distributions without the kinematic fit.
All three possible jet-boson associations are taken into account in the histograms.
\ref{fig:djm-csfit}a shows the dijet mass distribution of all Standard Model and SUSY point5 
events passing the selection cuts; \ref{fig:djm-csfit}b is the SM part of \ref{fig:djm-csfit}a; 
\ref{fig:djm-csfit}c and \ref{fig:djm-csfit}d are statistically independent template samples 
for $\tilde{\chi}^{\pm}_1$ and $\tilde{\chi}^0_2$, made by 500 fb$^{-1}$.
Before the fitting, the SM contribution (\ref{fig:djm-csfit}b) is subtracted from the 
distribution of all events (\ref{fig:djm-csfit}a). 
SUSY contributions other than $\tilde{\chi}^{\pm}_1$ and $\tilde{\chi}^0_2$ pair are
not corrected for, but the contribution is negligibly small.

Figure~\ref{fig:djm-csfit}e shows the result of a fit using a linear combination of the 
Chargino and Neutralino template distributions depicted \ref{fig:djm-csfit}c and d in. 
The residuals of the fit are displayed in figure~\ref{fig:djm-csfit}f. They are sufficiently 
small and don't show any specific structures, indicating a well working fit.


While it can be assumed that the SM distribution is well known and can be controlled for 
instance with the SM selection above, the assumption that the shape of the Chargino and 
Neutralino spectra is known is not evident. However, the shape of the dijet mass 
distribution on generator level is quite independent of the details of the SUSY scenario, 
as long as the decay into real $W$ and $Z$ bosons is open. As discussed  already in 
section~\ref{sec:smcontrol}, the shape of the reconstructed dijet mass distribution is 
influenced by the mass differences between $\tilde{\chi}^{\pm}_1$ / $\tilde{\chi}^{0}_2$ 
and the LSP, which determines the boost of the vector bosons and thus has an effect on the 
amount of combinatorical background and the mass resolution. As shown in the next section, 
the masses of the gauginos can be measured purely from edge positions in the energy spectra 
of the gauge bosons, without any assumption on the cross-section. Thus, with the gaugino 
masses measured, we are confident that enough is known about the SUSY scenario at hand 
to apply the template method.

The background subtraction and the fit have been performed 10000 times, varying the bin contents 
of the SUSY and the SM distribution according to their statistical errors. The fitted 
fractions of Chargino and Neutralino contribution have been averaged over all fit outcomes, 
while the expected uncertainty is estimated from the variance of the fit results. Expressed in 
percent of the expected cross-section, this procedure yields $99.97 \pm 0.84$\% for the Chargino 
and $97.50 \pm 2.75$\% for the Neutralino case.
In terms of absolute cross-sections this is equivalent to 
$\sigma(e^+e^- \rightarrow \tilde{\chi}^{+}_1\tilde{\chi}^{-}_1) = 124.80 \pm 1.05 $fb$^{-1}$ (MC: 124.84 fb$^{-1}$),
and $\sigma(\sigma(e^+e^- \rightarrow \tilde{\chi}^{0}_2\tilde{\chi}^{0}_2) = 21.90 \pm 0.62 $fb$^{-1}$ (MC: 22.46 fb$^{-1}$).

If we use a best jet pairing rather than all combinations for the dijet mass,
the statistical error grows by about 10\%. This illustrates the fact that the true 
jet-boson association cannot always be found and that the jet pairings not 
classified as ``best'' still contain valuable information.


%
%
\section{Mass Measurement}
\label{sect:mass}
The masses of gauginos can be obtained via the energy spectrum of the $W$ and $Z$ boson candidates, 
since the distribution of gauginos is box-like with edges determined by the masses 
and the center-of-mass energy. Deviations from the pure box shape are due to the finite width 
of the $W$ and $Z$ bosons, the beam energy spectrum and the detector resolution.
For the mass measurement, we have to separate the sample on an event-by-events basis 
into $\tilde{\chi}^{\pm}_1$ and $\tilde{\chi}^{0}_2$ pair candidates. This is done
via the dijet masses, as described in the next subsection. Afterwards, the edge positions 
are fitted for both the Chargino and Neutralino selected sample. Finally, the actual masses are
calculated from the edge positions.


\subsection{Dijet Selection}




For each event, the jet pairing with the highest probability in the kinematic fit is chosen. An event is selected as a Chargino or Neutralino candidate using the following $\chi^2$ variables, which are constructed from the invariant masses calculated from the four-vectors before the kinematic fit:

\begin{eqnarray}
	\chi^2_W(m_1, m_2) &=& \frac{(m_1 - m_W)^2 + (m_2 - m_W)^2}{\sigma^2} \\
	\chi^2_Z(m_1, m_2) &=& \frac{(m_1 - m_Z)^2 + (m_2 - m_Z)^2}{\sigma^2},
\end{eqnarray}

where $m_1$ and $m_2$ are dijet masses of selected jet-pairs,
$m_W$ and $m_Z$ are the nominal $W$ and $Z$ pole masses 
and $\sigma$= 5~GeV.
Events with $\chi^2_W < 4$ are classified as $\tilde{\chi}^{\pm}$, while events with $\chi^2_W > 4$ \& $\chi^2_Z < 4$ are selected as  $\tilde{\chi}^{0}_2$.

\begin{figure}[p] 
\begin{center}
\epsfig{figure=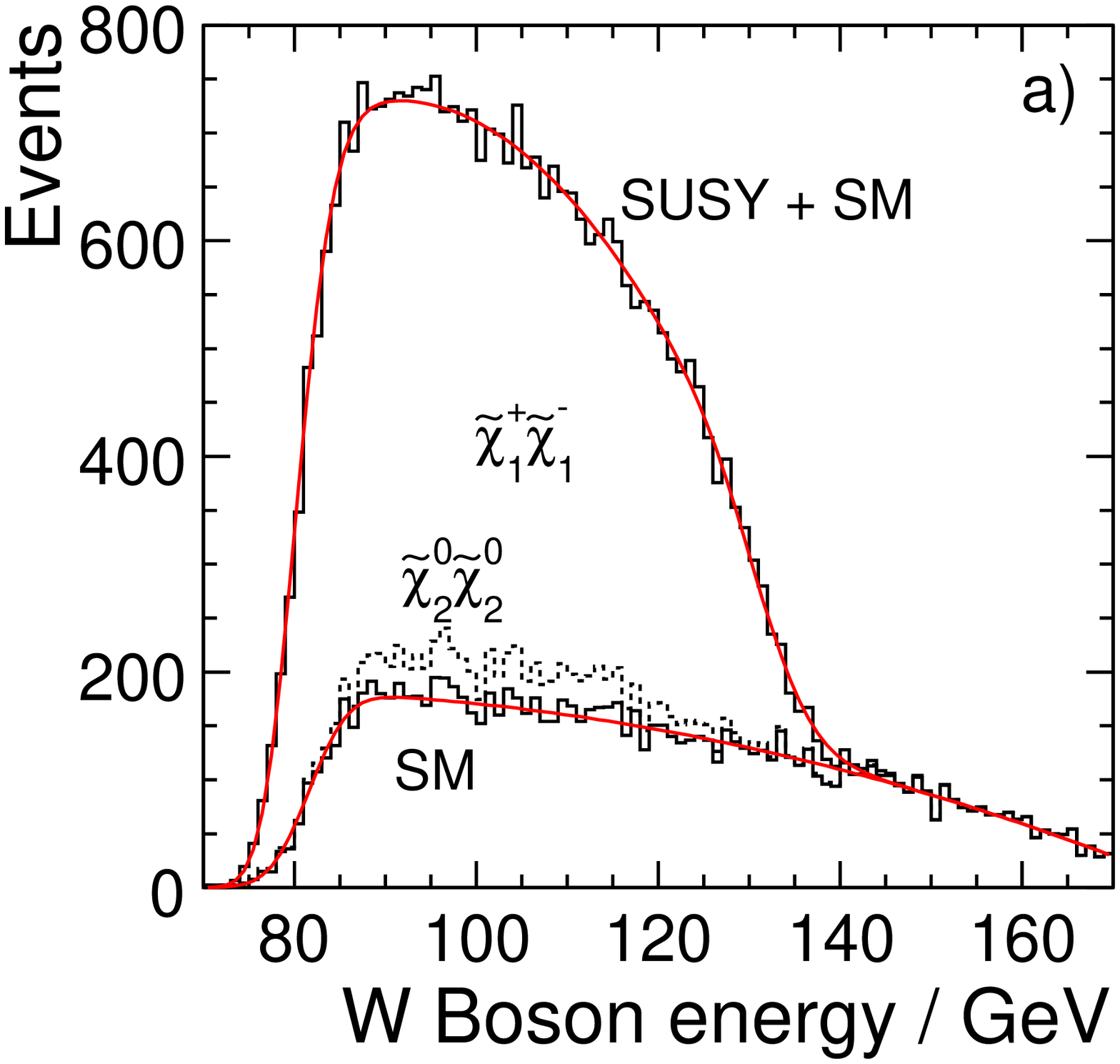,width=7cm}
\hspace{0.1cm}
\epsfig{figure=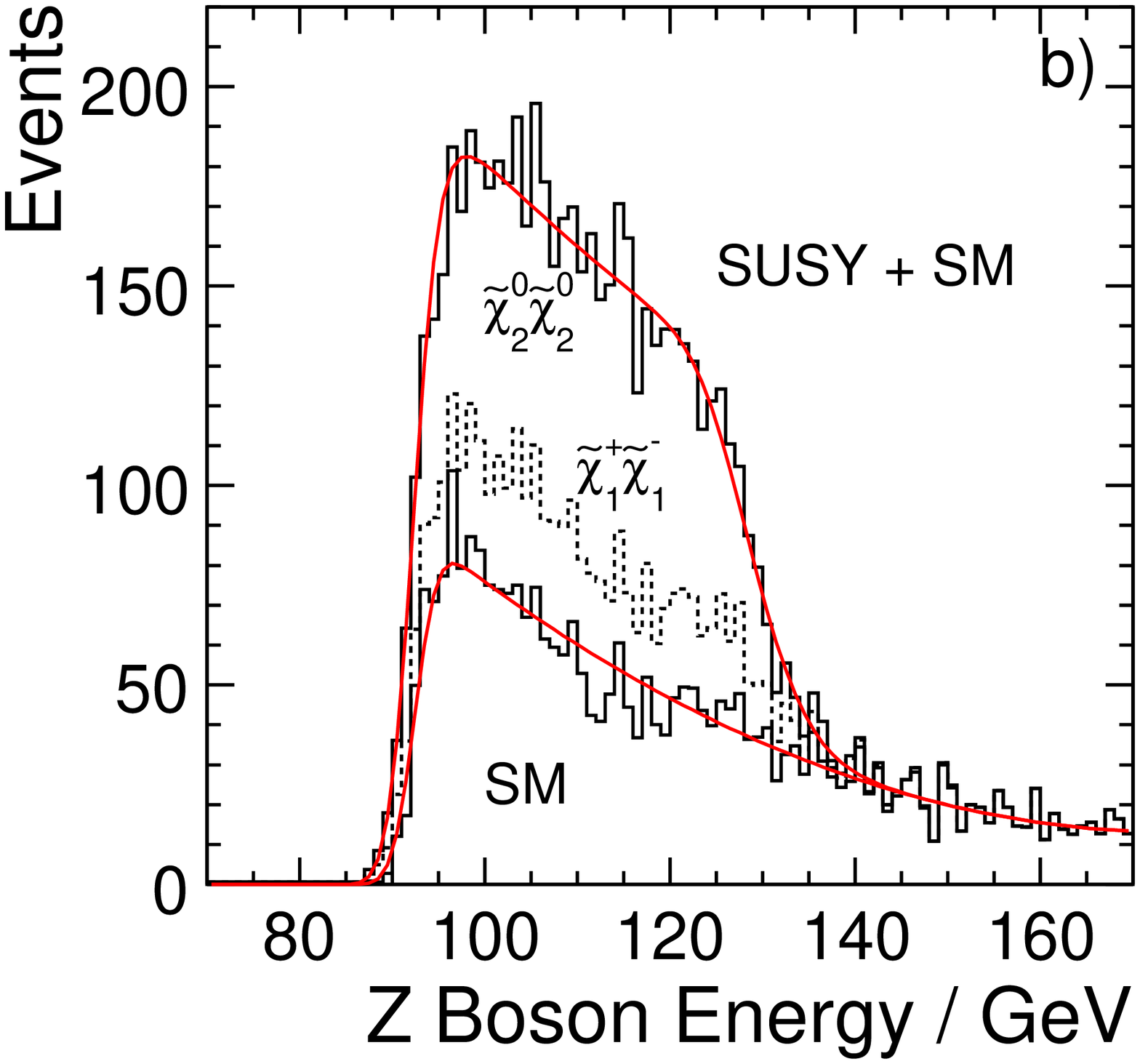,width=7cm}
\end{center}
	\caption{Mass determination: 
        a) Energy spectrum of the $W^{\pm}$ candidates reconstructed from events selected as $\tilde{\chi}^{\pm}_1$ pairs and 
        b) Energy spectrum of the $Z^0$ candidates reconstructed from events selected as $\tilde{\chi}^{0}_2$ pairs.
        In both cases, the Standard Model contribution has been fitted seperately before fitting the total spectrum.}
	\label{fig:mass2-kinfit}
\end{figure}

Figure~\ref{fig:mass2-kinfit}a) shows the energy spectrum of the selected $W$ candidates, while figure~\ref{fig:mass2-kinfit}b) presents the same spectrum for the $Z$ candidates.
The edge positions can be seen in the spectra, although the four-fermion background is still large, especially in the $Z$ energy distribution. The SM background can be fitted separately, as described below.

\subsection{Fitting the Edges}
\label{sec:fitfun}
In the next step, the energy spectra of the $W$ and $Z$ candidates are fitted according to
 the following procedure.
\begin{enumerate}
	\item First, the Standard Model contribution is fitted with the following function:
	\begin{equation}
		f_{SM}(x;t_0,a_{0-2},\sigma,\Gamma) = \int^\infty_{t_0}(a_2t^2+a_1t+a_0)V(t-x,\sigma,\Gamma)dt
	\end{equation}
	Here, $x$ denotes the boson energy, and $V(x,\sigma,\Gamma)$ is the Voigt function, 
        i.e. a Breit-Wigner function of width $\Gamma$ convoluted with a Gaussian of 
        resolution $\sigma$. The $t_0$ parameter adjusts the threshold position, while the parameters
        $a_0$, $a_1$ and $a_2$ are used to describe the shape of the plateau with a second order
        polynomial. The result of this fit is shown in figure~\ref{fig:mass2-kinfit}.
	\item Since the available statistics of the Standard Model sample is limited, the actual 
        background used in the SUSY fit is generated from the fitted functions, including 
        fluctuations according to the statistical errors expected from 500~fb$^{-1}$ of 
        integrated luminosity.  
	\item Finally, the sum of the SUSY spectra and the SM spectra generated in the 
         previous step 
         are fitted. The SUSY part of the fitting function is similar to the one used on the
         Standard Model, but this time also an upper edge position $t_1$ is introduced. 
         Furthermore, the Gaussian resolution $\sigma$ is allowed to have two different 
         values at the  
         edge positions, namely $\sigma_0$ and $\sigma_1$, with intermediate values obtained
         by linear interpolation.
	\begin{eqnarray}
		f(x;t_{0-1},b_{0-2},\sigma_{0-1},\Gamma) &=& f_{SM}+\int^{t_1}_{t_0}(b_2t^2+b_1t+b_0)V(t-x,\sigma(t),\Gamma)dt \\
		\sigma(t;\sigma_0,\sigma_1) &=& \sigma_0 + \frac{(\sigma_1-\sigma_0)(t-80)}{40}.
	\end{eqnarray}
	All parameters of $f_{SM}$ are fixed to the values obtained in the first step.
	For the $\tilde{\chi}^{0}_2$ fit, $b_2$ is also fixed to 0.
\end{enumerate}

Figure~\ref{fig:mass2-kinfit} shows the results of the SM fit as well as the 
results of SUSY mass fit for both the Chargino and the Neutralino selection.

To obtain edge positions, the fit is performed 100 times with different Standard Model 
spectra generated from the SM fit function. As final result, the averaged edge position and error
are given:

\begin{itemize}
\item $\tilde{\chi}^{\pm}_1$ lower edge: $79.88 \pm 0.19$ (MC: 79.80) GeV,
\item $\tilde{\chi}^{\pm}_1$ upper edge: $131.49 \pm 0.74$ (MC: 132.77) GeV,
\item $\tilde{\chi}^{0}_2$ lower edge: $92.34 \pm 0.44$ (MC: 93.09) GeV, and
\item $\tilde{\chi}^{0}_2$ upper edge: $127.67 \pm 0.76$ (MC: 129.92) GeV.
\end{itemize}

There is a tendency that the fitted numbers are slightly smaller than MC numbers.
Better jet energy correction or modification of the fitting function can reduce the shift,
but principally the shift could be corrected with a dedicated MC study.

\subsection{Mass Determination from Edge Positions}


The relation between the gaugino masses and the energy endpoints of the gauge bosons
is determined by pure kinematics. Neglecting radiation losses, the energy of the gauginos
is equal to the beam energy:$E_{\chi} = E_{\mathrm{beam}}$. In the gaugino restsystem, denoted with
$^\ast$, the energy of the vector boson (i.e. $W$ or $Z$) is given by the usual formula for 
two-body decays:
\begin{equation}
E_V^{\ast} = \frac{M_{\chi}^2 + M_V^2 - M_{\mathrm{LSP}}^2}{2 \cdot M_{\chi}},
\end{equation}
where subscript $\chi$ denotes the decaying gaugino (i.e. $\tilde{\chi}^{\pm}_1$ or 
$\tilde{\chi}^{0}_2$), $V$ the vector boson (i.e. $W$ or $Z$) and
the LSP $\tilde{\chi}^{0}_1$. Boosting this into the laboratory system yields:
\begin{equation}
E_V = \gamma E_V^{\ast} \pm \gamma \beta \sqrt{E_V^{\ast 2} - M_V^2}
\label{form:edges}
\end{equation}
The Lorentz boost $\gamma$ is given by $\gamma = E_{\chi}/M_{\chi}$, and 
$\beta = \sqrt{1-1/\gamma^2}$. The plus sign will give the upper edge of the allowed energy range,
$E_+$, and the minus sign the lower one, $E_-$. For further calculations it is useful 
to introduce the center point of the allowed energy range, $E_M$, and its width $E_D$:
\begin{equation}
E_M = \frac{E_+ + E_-}{2},   \quad E_D = \frac{E_+ - E_-}{2}
\end{equation}
In solving equation~\ref{form:edges} for the gaugino masses, it is useful to note that
$\gamma \cdot E_V^{\ast} = E_M$. With this relation, $E_V^{\ast}$ can be eliminated and 
thus the LSP mass in obtained from $E_D$:
\begin{eqnarray}
E_D &=& \gamma \sqrt{1-1/\gamma^2} \sqrt{E_V^{\ast 2} - M_V^2}\\
    &=&        \sqrt{1-1/\gamma^2} \sqrt{\gamma^2 \cdot E_V^{\ast 2} - \gamma^2 \cdot M_V^2}\\
    &=&        \sqrt{1-1/\gamma^2} \sqrt{E_M^2 - \gamma^2 \cdot M_V^2}
\end{eqnarray}
This is a quadratic equation in $\gamma^2$, which has two solutions:
\begin{equation}
\gamma^2 = \frac{1}{2 \cdot M_V^2} \left[ (E_+ \cdot E_- + M_V^2) \pm \sqrt{(E_+^2 - M_V^2)(E_-^2 - M_V^2)} \right]
\label{form:gamma2}
\end{equation}
Inserting this into $\gamma \cdot E_V^{\ast} = E_M$, the LSP mass can be solved for:
\begin{equation}
M_{\mathrm{LSP}}^2 = M_V^2 + \frac{E_{\mathrm{beam}}^2}{\gamma^2} \left(1 - \frac{E_+ + E_-}{E_{\mathrm{beam}}}\right)
\end{equation}
For a single energy spectrum, we thus have two solutions in the general case. However with the 
constraint that the LSP mass has to be the same for both the Chargino and the Neutralino decay,
a unique solution can be determined - in this case the one with the upper sign.


For the point5 SUSY parameters, the lower edge of the $W$ energy spectrum is just equal to the 
$W$ rest mass, meaning that the $W$ bosons from the decay can be produced at rest, with the LSP
carrying away all the momentum. This case has to be distinguished from a configuration where
the boost is so large that the $W$ could actually fly into the same direction as the LSP in the
laboratory frame. In this case, since the energy cannot become lower than the $W$ rest mass, 
the lower part of the spectrum would be ``folded over'' and create a second falling edge above 
the $W$ mass, precisely at $E_V = \sqrt{M_V^2 + p_{V,\mathrm{min}}^2}$, where 
$p_{V,\mathrm{min}} = -\gamma \beta E_V^{\ast} + \gamma \sqrt{E_V^{\ast 2} - M_V^2}$.
Moreover, this case of $E_- = M_W$ corresponds to the case where the equation for $\gamma^2$ has 
only one solution, with the $\pm$ term of equation~\ref{form:gamma2} vanishing. At this point,
the partial derivative $\partial E_-/\partial M_{\tilde{\chi}^{\pm}_1}$ becomes zero. So the 
inverse derivative which appears in the error propagation becomes undefined - or more realistically,
with $E_- = M_W$ not exactly fulfilled, at least very large.

Since the discrimination between models is beyond the scope of this paper, but will be subject of
future studies, we ignore here possible information from the lower edge of the $W$ energy spectrum. Instead, the lower and upper edge of the $Z$ energy spectrum are used to calculate the
masses of $\tilde{\chi}^{0}_2$ and $\tilde{\chi}^{0}_1$. In a second step, the Chargino mass is 
calculated from the LSP mass and the upper edge of the $W$ spectrum.  


The error propagation is done by using a toy Monte Carlo, taking into account the correlations between the two masses determined from one energy spectrum. It calculates the gaugino masses by above equations with edge positions varying randomly according to their errors obtained from the edge fit. For the center edge positions two patterns were tried, the fitted edge positions 
 and the MC truth positions. 

Table~\ref{tab:masses} shows the obtained mass values and errors. 
Without correction of the edge position, the average value of obtained masses deviates by 3-4 GeV 
from the MC truth. This might be due to the fact that phase space was not considered, and could be 
reduced by an improved fitting function.
with better fitting functions. Without the kinematic fit, the mass resolution is worse by 
typically 400 to 500~MeV, which corresponds to 15 to 40\% of the errors, depending on the 
gaugino considered.

\section{Summary}
\label{sect:summ}
The physics performance of the ILD detector concept has been evaluated using a SUSY benchmark 
scenario referred to as ``Point 5'', where $\tilde{\chi}^{\pm}_1$ and $\tilde{\chi}^{0}_2$ 
are nearly mass degenerate and decay into real $W^{\pm}$ and $Z^0$ bosons, respectively, 
plus a $\tilde{\chi}^{0}_1$. The cross-sections for Chargino and Neutralino pair production 
have been obtained by a fit to the two-dimensional dijet mass spectrum relying on Monte-Carlo 
templates. The resulting statistical errors are 0.84\% in the Chargino case and 2.75\% in the 
Neutralino case.

The gaugino masses have been determined from a fit to the edges of the energy spectra of the 
$W^{\pm}$ and $Z^0$ bosons obtained by a kinematic fit.
The resulting mass resolutions are 2.9~GeV, 1.7~GeV and 1.0~GeV for $\tilde{\chi}^{\pm}_1$, 
$\tilde{\chi}^{0}_2$ and $\tilde{\chi}^{0}_1$, respectively. Without the kinematic fit, 
the mass resolution is worse by 400 to 500~MeV.

\section*{Acknowledgements}
We thank Frank Gaede, Steve Aplin, Jan Engels and Ivan Marchesini for simulating the event samples for this study, and we thank Timothy Barklow and Mikael Berggren for generating the corresponding four-vector files for SM backgrounds and SUSY, respectively. We further thank Fran\c{c}ois Richard and Tohru Takeshita for the fruitful discussions about the analysis. This work has been supported by the Emmy-Noether programme of the Deutsche Forschungsgemeinschaft (grant LI-1560/1-1).

\newpage

%
%

\begin{table}
	\begin{center}
		\scriptsize{
		\begin{tabular}{|l||r|r|r|r|r|r|r|} \hline
                                      & $\tilde{\chi}^{+}_1\tilde{\chi}^{-}_1 \rightarrow$ hadrons &$\tilde{\chi}^{0}_2\tilde{\chi}^{0}_2 \rightarrow$ hadrons& other SUSY& SM $\gamma\gamma$    & SM 6f   & SM 4f         & SM 2f   \\ \hline\hline
nocut                                 & 28529  & 5488  & 74650   & 3.66e+09 & 521610  & 1.48e+07 & 2.14e+07 \\ \hline
Total \# of tracks $\geq 20$          & 27897  & 5449  & 24305   & 3.03e+06 & 495605  & 6.68e+06 & 5.33e+06 \\ \hline
$100<E_\mathrm{vis}<300$ GeV          & 27895  & 5449  & 22508   & 1.06e+06 & 44394   & 959805   & 1.56e+06 \\ \hline
$E_\mathrm{jet}>5$		      & 27889  & 5446  & 20721   & 908492   & 44096   & 916507   & 1.47e+06 \\ \hline
$|\cos(\theta)_\mathrm{jets}|<0.99$   & 26560  & 5240  & 19200   & 350364   & 41098   & 678083   & 874907   \\ \hline
$y_{34}>0.001$			      & 26416  & 5218  & 15255   & 202510   & 38638   & 423080   & 166305   \\ \hline
\# of tracks $\geq 2 / $jets	      & 25717  & 5146  & 9559    & 162193   & 22740   & 255870   & 145270   \\ \hline
$|\cos{\theta_{\mathrm{miss}}}|<0.99$ & 25463  & 5099  & 9487    & 25087    & 22311   & 193706   & 4039     \\ \hline
$E_\mathrm{l}<25$	              & 25123  & 4981  & 6463    & 23133    & 14407   & 154927   & 3534     \\ \hline\hline
$N_{\mathrm{PFO}}> 3$                 & 25029  & 4975  & 6103    & 23014    & 13696   & 139429   & 3518     \\ \hline
$|\cos{\theta_{\mathrm{miss}}}|<0.8$  & 20144  & 4079  & 5180    &   681    &  9950   &  62668   &  529     \\ \hline
$M_{\mathrm{miss}} > 220$~GeV         & 20139  & 4079  & 5180    &   630    &  3687   &  45867   &  389     \\ \hline
kin. fit converged                    & 20085  & 4068  & 4999    &   626    &  3649   &  44577   &  341     \\ \hline
		\end{tabular}
		}
	\caption{Event numbers after each of the selection cuts, normalized to 500 fb$^{-1}$ and $P(e^+,e^-) = (30\%, -80\%)$.}
	\label{tbl:cutstat}
	\end{center}
\end{table}

\begin{table}
	\begin{center}
		\begin{tabular}{|l||r|r|r|r|} \hline
Processes                                              & No cut   & all cuts & Purity & Efficiency           \\ \hline\hline
$\tilde{\chi}^+_1\tilde{\chi}^-_1 \rightarrow$ hadrons & 28529    & 16552    & 58\%   & 58\%                 \\ \hline
$\tilde{\chi}^0_2\tilde{\chi}^0_2 \rightarrow$ hadrons & 5488     &  3607    & 13\%   & 65\%                 \\ \hline
Other SUSY point5                                      & 74650    &    77    & 0.27\% & $1.0 \times 10^{-3}$ \\ \hline\hline
qqqq (WW, ZZ)                                          & 4.29e+06 &  5885    & 21\%   & $1.4 \times 10^{-3}$ \\ \hline
qq$\ell\nu$ (WW)                                       & 5.19e+06 &   561    & 2.0\%  & $1.1 \times 10^{-4}$ \\ \hline
qqqq$\ell\nu$ (tt)                                     & 216996   &   489    & 1.7\%  & $2.3 \times 10^{-3}$ \\ \hline
$\gamma\gamma\rightarrow$qqqq                          & 26356    &   397    & 1.4\%  &  1.5\%               \\ \hline
qqqq$\nu\nu$ (WWZ)                                     & 9262     &   268    & 0.94\% &  2.9\%               \\ \hline
qq$\nu\nu$ (ZZ)                                        & 367779   &    76    & 0.27\% & $2.1 \times 10^{-4}$ \\ \hline
qq                                                     & 9.77e+06 &    76    & 0.27\% & $7.8 \times 10^{-6}$ \\ \hline
Other background                                       & 3.68e+09 &   438    & 1.5\%  & $1.2 \times 10^{-7}$ \\ \hline
		\end{tabular}

	\caption{Purity and efficiency of signal and major background sources after the selection cuts 
                 and with an invariant dijet mass larger than 65 GeV.
The processes in pathentheses indicate the dominant intermediate states.}
	\label{tbl:cutstat2}
	\end{center}
\end{table}

\begin{table}
	\begin{center}
	\begin{tabular}{|l|r|r|r|} \hline
Observables                & Obtained value & Error	 & Error at the true mass \\\hline\hline
$m(\tilde{\chi}^{\pm}_1)$  &  220.90 GeV    & 2.90 GeV   & 3.34 GeV \\\hline
$m(\tilde{\chi}^{0}_2)$    &  220.56 GeV    & 1.72 GeV   & 1.39 GeV \\\hline
$m(\tilde{\chi}^{0}_1)$    &  118.97 GeV    & 1.02 GeV   & 0.95 GeV \\\hline
	\end{tabular}
	\caption{Performance on gaugino masses and associated errors. The last column shows 
        errors on masses
	when the true edge positions are used in the error propagation. MC truth masses are 
        216.7, 216.5 and 115.7 GeV for
	$\tilde{\chi}^{\pm}_1$, $\tilde{\chi}^{0}_2$ and $\tilde{\chi}^{0}_1$, respectively.}
	\label{tab:masses}
	\end{center}
\end{table}

\end{document}